\documentclass[prl,twocolumn,superscriptaddress]{revtex4-1} 
\usepackage{color}
\usepackage[colorlinks=true,linkcolor=blue,citecolor=blue,urlcolor = blue]{hyperref}
\usepackage{mathtools}
\usepackage{amsmath} 
\usepackage{graphicx} 
\usepackage{lineno} 
\usepackage{braket}

\usepackage{hyperref}

\begin{document}

\begin{abstract}
We perform precise studies of two- and three-body interactions near an intermediate-strength Feshbach resonance in $\prescript{39}{}{\mathrm{K}}$ at $33.5820(14)\thinspace$G. Precise measurement of dimer binding energies, spanning three orders of magnitude, enables the construction of a complete two-body coupled-channel model for determination of the scattering lengths with an unprecedented low uncertainty. Utilizing an accurate scattering length map, we measure the precise location of the Efimov ground state to test van der Waals universality. Precise control of the sample's temperature and density ensures that systematic effects on the Efimov trimer state are well understood. We measure the ground Efimov resonance location to be at $-14.05(17)$ times the van der Waals length $r_{\mathrm{vdW}}$, significantly deviating from the value of $-9.7 \thinspace r_{\mathrm{vdW}}$ predicted by van der Waals universality. We find that a refined multichannel three-body model, built on our measurement of two-body physics, can account for this difference and even successfully predict the Efimov inelasticity parameter $\eta$.
\end{abstract}

\title{Precision Test of the Limits to Universality in Few-Body Physics}

\author{Roman Chapurin}
\affiliation{JILA, National Institute of Standards and Technology, and the University of Colorado, Department of Physics, Boulder, Colorado 80309, USA}
\author{Xin Xie}
\affiliation{JILA, National Institute of Standards and Technology, and the University of Colorado, Department of Physics, Boulder, Colorado 80309, USA}
\author{Michael J. Van de Graaff}
\affiliation{JILA, National Institute of Standards and Technology, and the University of Colorado, Department of Physics, Boulder, Colorado 80309, USA}
\author{Jared S. Popowski}
\affiliation{JILA, National Institute of Standards and Technology, and the University of Colorado, Department of Physics, Boulder, Colorado 80309, USA}
\author{Jos\'e P. D'Incao}
\affiliation{JILA, National Institute of Standards and Technology, and the University of Colorado, Department of Physics, Boulder, Colorado 80309, USA}
\author{Paul S. Julienne}
\affiliation{Joint Quantum Institute, National Institute of Standards and Technology, and the University of Maryland, College Park, Maryland 20742, USA}
\author{Jun Ye}
\affiliation{JILA, National Institute of Standards and Technology, and the University of Colorado, Department of Physics, Boulder, Colorado 80309, USA}
\author{Eric A. Cornell}
\affiliation{JILA, National Institute of Standards and Technology, and the University of Colorado, Department of Physics, Boulder, Colorado 80309, USA}

\date{\today}
\maketitle

The few- and many-body physics of an interacting gas are intractable when treated in full microscopic detail. However, the problem can be greatly simplified in a dilute ultracold atomic gas with near-resonant interactions, where the two-body scattering length $a$ greatly exceeds the van der Waals length $r_{\mathrm{vdW}}$ characterizing the range of the interacting potential. In such a scenario, all physical observables can be parametrized by only two dimensionless quantities describing the strength of interactions and the level of quantum degeneracy \cite{Braaten2006}: $n a^{3}$ and $n \lambda ^{3}$, where $n$ is the atomic density and $\lambda$ is the thermal wavelength. Then, continuous scaling transformations such as $n \rightarrow \zeta ^{-3} n$, $a \rightarrow \zeta a$, and $\lambda \rightarrow \zeta \lambda$ will leave all observables and their dynamics invariant when measured in rescaled units. Such behavior is regarded as universal, insensitive to microscopic details in the problem and the chosen atomic species. 

Nevertheless, the principle of universality has its limitations. For example, unless all length scales in the problem ($\left| a \right|$, $\lambda$, $n^{-1/3}$, etc.) greatly exceed $r_{\mathrm{vdW}}$, nonuniversal corrections due to short-ranged physics must be implemented. Even when these conditions are well satisfied, a more fundamental effect concerning few-body interactions can break universality: the Efimov effect~\cite{Efimov1970}. Within this phenomenon, short-ranged near-resonant two-body interactions give rise to a three-body attraction that hosts an infinite series of Efimov trimer states. Each consecutive state meets the three-body continuum at a particular value of scattering length that is 22.7 times larger than the previous state, with $a_{-}$ defining the ground state location \cite{Braaten2006}. Although these fixed length scales break the continuous aspect of universality, there remains a discrete version of scale transformations, with $\zeta$ values restricted to $22.7^{j}$, where $j$ is an integer. 

The value of $a_{-}$ was originally thought to be set by the details of the short-range interaction, and therefore to be thoroughly nonuniversal. However, it was noted that across many atomic species and different Feshbach resonances the measured $a_{-}$ value was within $20\%$ of $-9 \thinspace r_{\mathrm{vdW}}$ \cite{Berninger2011, Naidon2017, Greene2017, DIncao2018}. This suggested that $a_{-}$ depends only on the longest-range part of the short-range physical interaction. Theory indeed predicts a similar value of $a_{-} = - 9.7 \thinspace r_{\mathrm{vdW}}$ \cite{Wang2012, naidon2014PRL, Naidon2012, Naidon2017, Greene2017, DIncao2018, chin2011arxiv}. This ``van der Waals universality,'' together with the Efimov scaling, allows one to predict the full Efimov structure to arbitrarily large length scales.

\begin{figure}
\includegraphics[width=8.6cm,keepaspectratio]{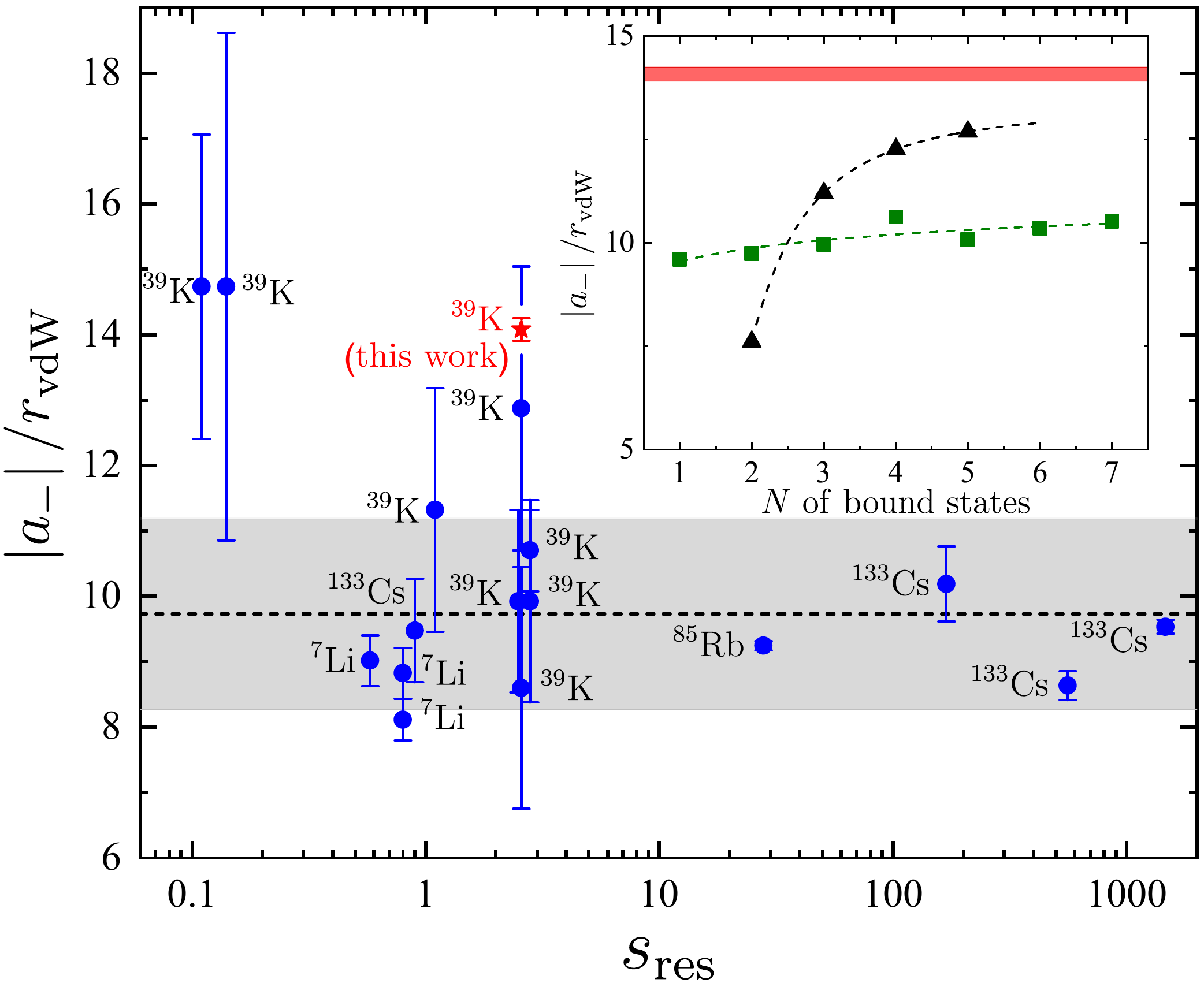} 
\caption{Survey of experimental $a_{-}$ values in homonuclear systems, inspired by \cite{Johansen2017}. Previous results (blue circles) \cite{Zaccanti2009, Gross2011, Berninger2011, wild2012PRL, Roy2013, Dyke2013, Huang2014,Wacker2018} show a tentative dependence of the $a_{-}$ value on the Feshbach resonance strength parameter $s_{\mathrm{res}}$. Our measurement (red star; red band in the inset) is the strongest evidence of departure from the $-9.7 \pm 15\% \thinspace r_{\mathrm{vdW}}$ value (dashed line and gray area) predicted by van der Waals universality \cite{Wang2012, Naidon2012, Naidon2017, Greene2017}. The inset shows calculations for $a_{-}$ based on a single van der Waals potential \cite{Wang2012} with $N=1\text{--}7$ $s$-wave two-body bound states (green squares) and results from our multichannel model \cite{SupplemMat} with $N=2\text{--}5$ (black triangles).}\label{fig:aminus_vs_sres} 
\end{figure}

Our experimental goal is to definitively challenge the robustness of this van der Waals universality. It has been speculated \cite{Petrov2004, Gogolin2008, Schmidt2012, Wang2014, Johansen2017, Langmack2018, Roy2013, Note1}\footnotetext[1]{The term van der Waals universality is used in a rather different, more inclusive multichannel context in Ref.~\cite{Wang2014}} that universality of the Efimov structure depends on the breadth of the Feshbach resonance quantified by the dimensionless parameter $s_{\mathrm{res}}$. Very roughly, $s_{\mathrm{res}}$ may be understood as the parameter that characterizes the range of the scattering length, $|a| \gtrsim 4 r_{\mathrm{vdW}}/s_{\mathrm{res}}$, over which the two-body Feshbach resonance has universal structure, meaning, e.g., that the two-body binding energy is $E_{b} = \hbar^{2}/(m a^{2})$ \cite{sresSupplemMat}. One might expect the three-body Efimov resonances to be more precisely universal when they fall more deeply into that range of $a$ for which the two-body Feshbach resonant structure is universal. In previous experiments on homonuclear Efimov states \cite{Zaccanti2009, Pollack2009, Gross2009, Gross2010, Gross2011, Berninger2011, wild2012PRL, Roy2013, Dyke2013, Huang2014, Wacker2018}, there is some support for the notion that, as $s_{\mathrm{res}}$ gets smaller, the measured $a_{-}$ values should begin to deviate from the universal $a_{-} = - 9.7 \thinspace r_{\mathrm{vdW}}$ value; see Fig.~\ref{fig:aminus_vs_sres}. However, this conclusion is only tentative due to large experimental uncertainties in the measured $a_{-}$ \cite{Roy2013}, unexpected temperature dependence \cite{Wacker2018}, and large systematic uncertainties in the parameters of the underlying two-body Feshbach resonance \cite{Zaccanti2009, Roy2013, Dyke2013, Pollack2009}. Although there are intriguing experimental \cite{barontini2009PRL, bloom2013PRL, hu2014PRA, pires2014PRL, tung2014PRL, maier2015PRL, ulmanis2016PRL, Johansen2017, williams2009PRL, lompe2010, nakajima2011PRL} and theoretical \cite{helfrich2010PRA, wang2013ultracold, wang2015PRL, ulmanis2016PRL} results for the heteronuclear cases, the possible influence of many additional parameters (mass ratio, quantum statistics, and inter- and intraspecies scattering lengths) makes the question of universality in those systems a topic for an entirely separate investigation.

In this Letter, we present a precise test of van der Waals universality near a Feshbach resonance with $s_{\mathrm{res}}=2.57$ \cite{SupplemMat}, which is intermediate between the narrow ($s_{\mathrm{res}} \ll 1$) and broad ($s_{\mathrm{res}} \gg 1$) regimes. Specifically, we accurately determine the value of $a_{-}$ by having precise control of critical experimental parameters such as temperature, density and scattering length. Because of our tight control of both systematic and statistical errors, ours is the first measurement of a compelling nonuniversal $a_{-}$ value in a homonuclear Efimov resonance. 

A thorough characterization of the Feshbach resonance and an accurate map of the scattering length are required for precise determination of the $a_{-}$ value. Accordingly, we perform high-precision spectroscopy on a pure gas of Feshbach dimers and accurately determine their binding energies. This measurement enables us to refine our two-body model and accurately predict the scattering length in our Efimov measurements \cite{SupplemMat}. In other Feshbach resonance studies, methods based on number loss or thermalization rate have occasionally given inconsistent results. By contrast, dissociation spectroscopy of Feshbach dimers isolates two-body physics and accurately determines resonance properties \cite{Chin2005, Zurn2013, Julienne2014, Shkedrov2018}. 

Precision molecular spectroscopy requires long interrogation times under unperturbed conditions. We stabilize the magnetic field to the milligauss-level and eject all unpaired atoms, whose presence affects dimer lifetimes and complicates the spectroscopy. A pure molecular sample is prepared by starting with $\sim 10^{5} \thinspace$atoms confined in an optical dipole trap and a temperature of $\sim 300 \thinspace$nK. We transfer a fraction of atoms in the $\ket{F=1,m_F=-1}$ hyperfine state to the dimer state by magnetoassociation \cite{Kohler2006}. Subsequently, all residual unpaired atoms are blasted away by multiple radio-frequency (rf) and optical pulses, leaving a pure sample of $\sim 10^{4} \thinspace$molecules. Lastly, the magnetic field $B$ is ramped to various values, corresponding to different binding energies, where we perform rf spectroscopy. 

\begin{figure}
\includegraphics[width=8.6cm,keepaspectratio]{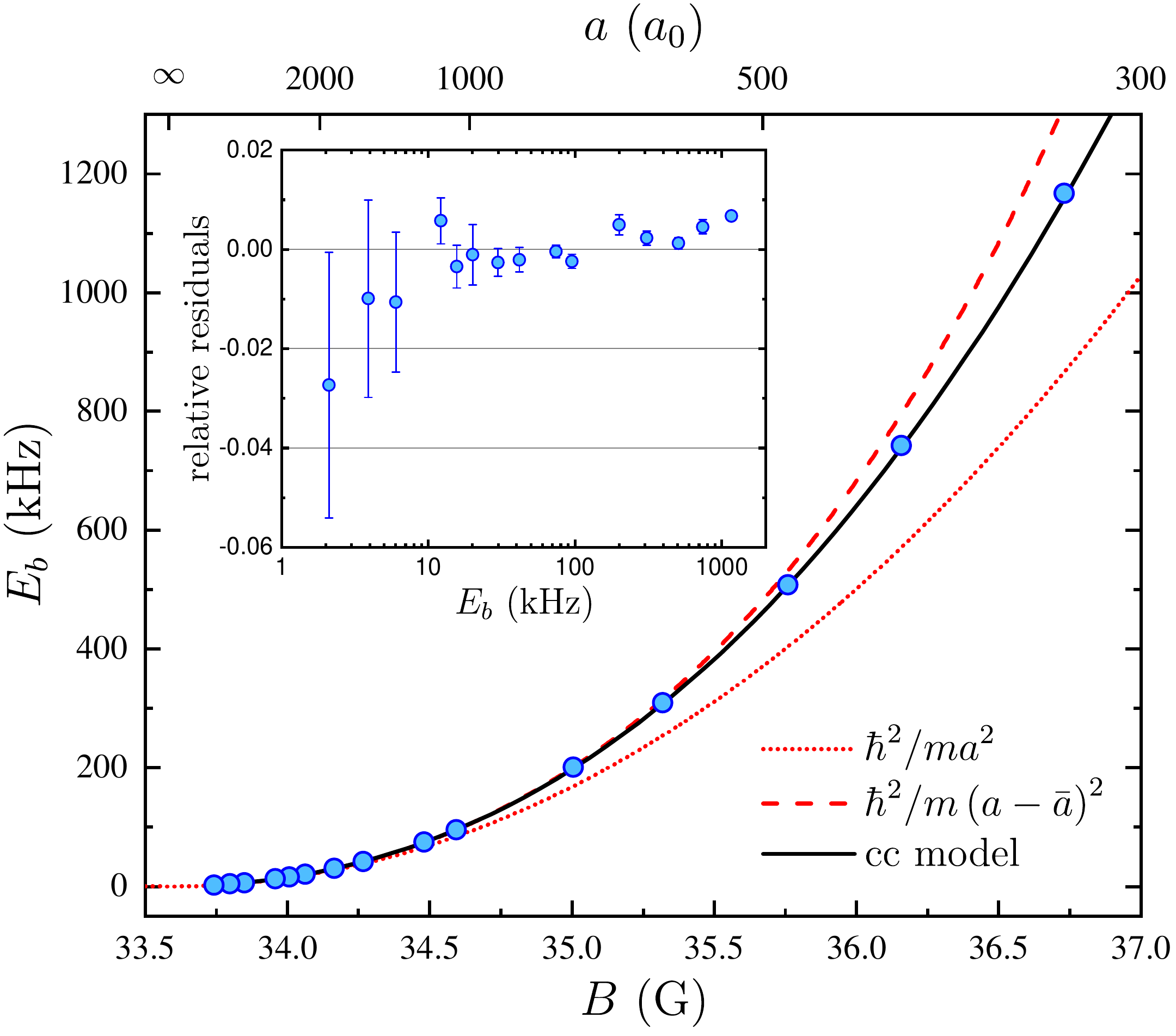} 
\caption{Precise measurement of Feshbach dimer binding energies $E_{b}$ as a function of magnetic field $B$. Small experimental uncertainties on $E_{b}$, spanning from $56\thinspace$Hz at $E_{b}/h=2.103 \thinspace$kHz to $1.0\thinspace$kHz at $E_{b}/h=1167.2 \thinspace$kHz, are not resolvable in the figure. A coupled-channel (cc) model is required to describe our data \cite{SupplemMat}. The solid curve shows the resulting fit and the inset shows remarkably small fractional residuals. Contrary to applicability near broad Feshbach resonances \cite{Chin2010}, universal expressions (dashed and dotted curves) are insufficient for describing $E_{b}$ near our intermediate strength resonance.
}
\label{fig:dimer_energy} 
\end{figure}

We dissociate molecules by transferring one atom of the pair from the $\ket{F=1,m_F=-1}$ interacting state to the $\ket{F=1,m_F=0}$ imaging state. The final state being nearly noninteracting enables us to directly probe the dimer binding energy. Additionally, the transition being magnetically less sensitive near $B$ values of interest allows long molecular interrogation times, limited only by dimer lifetimes, to achieve high spectral resolution. We scan the rf frequency and measure the transferred fraction, keeping the pulse energy low to limit saturation effects and dissociate a maximum $50\%$ of molecules. We fit the measured spectrum to a functional form given by the Franck-Condon factor of the bound-free transition \cite{Chin2005}, and we extract the molecular binding energy $E_{b}$ \cite{SupplemMat,Idziaszek2006}. We repeat this procedure to determine $E_{b}$ at different magnetic field values, as depicted in Fig.~\ref{fig:dimer_energy}. 

The universal expression $E_{b} = \hbar^{2}/(m a^{2})$ is always accurate for large enough $a$. A more refined expression $E_{b} = \hbar^{2}/[m \left(a - \bar{a}\right)^{2}]$, which introduces the mean scattering length $\bar{a} \approx 0.956 \thinspace r_{\mathrm{vdW}}$ \cite{Gribakin1993}, is valid at smaller values of $a$ as long as $a\gg r_{\mathrm{vdW}} / s_{\mathrm{res}}$ \cite{Chin2010}. However, such treatments are inadequate for narrow and intermediate resonances. To better compare to our experimental data, we developed a coupled-channel model \cite{SupplemMat} capable of describing our high-precision $E_{b}$ data. We fine-tune the model's parameters, the singlet and triplet scattering potentials, to accurately match most of our measurements to within $1\%$, as depicted in Fig.~\ref{fig:dimer_energy}'s inset. As a result, we determine a particular linear combination of the singlet and triplet scattering lengths $0.2470 \thinspace a_{S} + 0.9690 \thinspace a_{T} = 1.926(2) \thinspace a_{0}$ \cite{SupplemMat}, further constraining the previously reported values of $a_{S} = 138.49(12)\thinspace a_{0}$ and $a_{T} = -33.48(18)\thinspace a_{0}$ \cite{Falke2008, DErrico2007}. Furthermore, we constrain the Feshbach resonance location to within $33.5820(14)\thinspace \mathrm{G}$ \cite{SupplemMat}, which is a two-orders-of-magnitude improvement over the previous measurement \cite{Roy2013} and an unprecedented \cite{Zurn2013} accuracy better than $3\times 10^{-5}$ of the resonance width. 

With a good grasp on two-body physics, we seek to test the validity of van der Waals universality near our \mbox{Feshbach} resonance. We perform precision atom-loss spectroscopy to obtain the Efimov ground state location $a_{-}$ \cite{Kraemer2006}. Specifically, we measure the inelastic three-body recombination coefficient $L_{3}$ in the vicinity of $a_{-}$, where the presence of the nearby Efimov state leads to a resonant enhancement of the three-body loss: an Efimov resonance. A zero-temperature zero-range expression \cite{Braaten2006} relates $L_{3}$ features to $a_{-}$ for $a<0$: 
\begin{align}
L_{3}^{T=0}(a) \approx \frac{3 \hbar a^{4}}{m} \frac{4590 \sinh(2\eta)}{\sin^{2}[s_{0}\ln(a/a_{-})]+\sinh^{2}(\eta)} \,,
\label{eq:L3_T0}
\end{align}
where the dimensionless inelasticity parameter $\eta$ characterizes the Efimov resonance width and the constant $s_{0} \approx 1.00624$ fixes Efimov series spacing $e^{\pi / s_{0}} \approx 22.7$. Although Eq.~(\ref{eq:L3_T0}) adequately describes $L_{3}$ in the limit of $\lambda \gg \left|a\right|$, for increasing temperatures, it becomes less valid; and a finite-temperature zero-range model \cite{Rem2013, Petrov2015} is required to describe the three-body loss for $a<0$:
\begin{align}
\label{eq:L3_T_dependent}
L_{3}(a, T) = &A \frac{72 \sqrt{3} \pi ^{2} \hbar \left( 1-e^{-4\eta} \right) }{m k_{\mathrm{th}}^{6} a^{2}} \\
& \times \int_{0}^{\infty} \frac{\left( 1 - \left| s_{11}(x) \right|^{2} \right) e^{-x^{2} / \left(k_{\mathrm{th}} a \right)^{2}} x}{\left| 1 + s_{11}(x) \left(\frac{-x a_{-}}{1.017\left|a\right|} \right)^{-2 i s_{0}} e^{-2 \eta} \right|^{2}} dx \notag \,,
\end{align}
where $k_{\mathrm{th}} = \sqrt{m k_{\mathrm{B}} T} / \hbar$, $x = k |a|$, $A$ is a numerical factor that improves the fit quality by allowing for uncertainty in the absolute density, and the complex function $s_{11}(x)$ is an $S$-matrix element from Refs.~\cite{Rem2013} and \cite{Petrov_s11}. 

We perform $L_{3}(a)$ measurements at different temperatures and extract $a_{-}$ using the zero-range model [Eq.~(\ref{eq:L3_T_dependent})]. We begin with dilute thermal samples at $a = -100 \thinspace a_{0}$. We ensure our gas is fully thermalized and make trapping potentials sufficiently deep to be certain that evaporative losses have a negligible effect on our measurements. Then, we ramp $a$ to a value of interest and let three-body loss occur for a varied amount of time, allowing up to $30\%$ decay of the initial atom number. Subsequently, we ramp $a$ to a value of $-200 \thinspace a_{0}$, transfer the remaining atoms to the $\ket{F=2,m_F=-2}$ state, and perform time-of-flight imaging. We determine the time-dependent density $n$ from the measured temperatures and atom numbers \cite{SupplemMat}. For each scattering length, we extract the $L_{3}$ value by numerically solving the expression \cite{Weber2003}:
\begin{align}
\label{eq:L3_loss}
\frac{1}{N} \frac{dN}{dt} = - L_{3} \langle n^{2} \rangle - \alpha \,,
\end{align}
where $\langle n^{2} \rangle = (1/N)\int{n^{3}(\vec{x}) d^{3}x}$ and the constant $1/\alpha > 40\thinspace$s is the $a$-independent one-body decay time measured at $a=-100 \thinspace a_{0}$, which is negligible as compared to the three-body loss timescales of $50 \text{--} 170 \thinspace$ms for our $n$ near $a_{-}$. Additionally, we check that the two-body loss contribution $-L_{2} \langle n \rangle$ to Eq.~(\ref{eq:L3_loss}), with $L_{2}$ predicted by our two-body model, is also negligible.

\begin{figure}
\includegraphics[width=8.6cm,keepaspectratio]{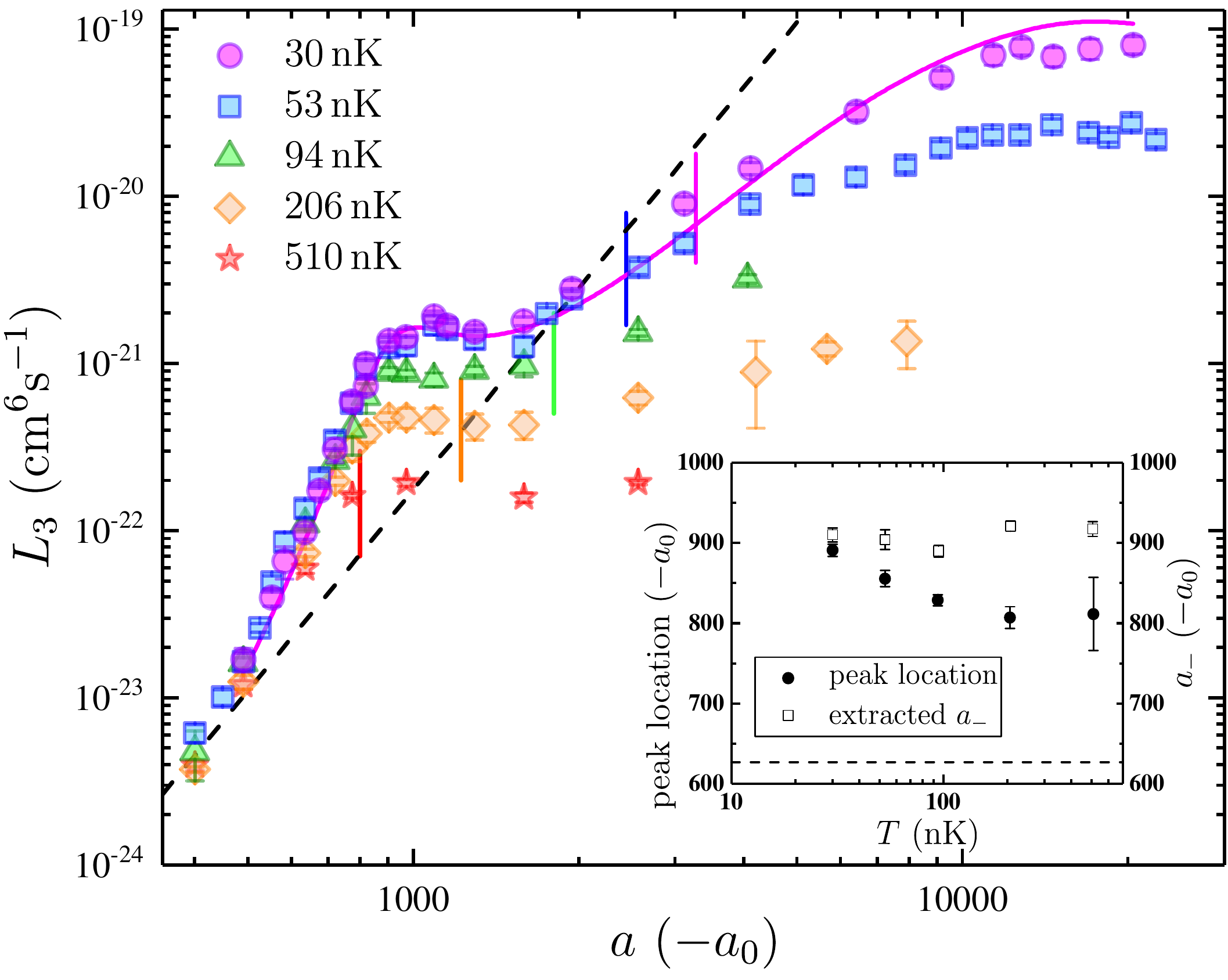} 
\caption{Temperature dependence of the three-body loss coefficient $L_{3}$, scaling as $a^{4}$ scaling (dashed) \cite{Fedichev1996, Esry1999, Weber2003}, enhanced near an Efimov ground state located at $a_{-}$. For each temperature, we fit our data using a zero-temperature zero-range model [Eq.~(\ref{eq:L3_T0})], limiting fits to data points for which $\left|a\right| < \lambda /10$ (short vertical lines), to extract the $L_{3}/a^{4}$ peak location and a finite-temperature zero-range model [Eq.~(\ref{eq:L3_T_dependent}), solid] to extract the true $a_{-}$ value. The inset shows the extracted peak locations (circles) and $a_{-}$ values (squares), where both coincide at the lowest temperature. The observed $a_{-}$ value significantly deviates from the $a_{-} = - 630 \thinspace a_{0}$ value (inset dashed line) predicted by van der Waals universality~\cite{Wang2012}. 
}
\label{fig:L3_vs_T} 
\end{figure}

Accurate calibration of $a$ and density (and not just relative changes) enables accurate comparison of the measured $L_{3}(a)$ values at different temperatures, as depicted in Fig.~\ref{fig:L3_vs_T}. We fit the data collected at each temperature to Eq.~(\ref{eq:L3_T_dependent}) with three parameters: $a_{-}$ (see inset of Fig.~\ref{fig:L3_vs_T}), $\eta$, and $A$. We take the weighted mean across all temperatures to extract single values $a_{-} = - 908(11) \thinspace a_{0} = -14.05(17)\,r_{\mathrm{vdW}}$ and $\eta = 0.25(1)$ \cite{SupplemMat}. Equation~(\ref{eq:L3_T_dependent}) will eventually become inaccurate at large $a$, if only because the functional form requires the first two resonances be a factor of 22.7 apart. We vary the fit range from all $a$ to only $\left|a\right| < \lambda / 10$ and take the maximal spread of all fit errors as the uncertainty on $a_{-}$ and $\eta$.

\begin{figure}
\includegraphics[width=8.6cm,keepaspectratio]{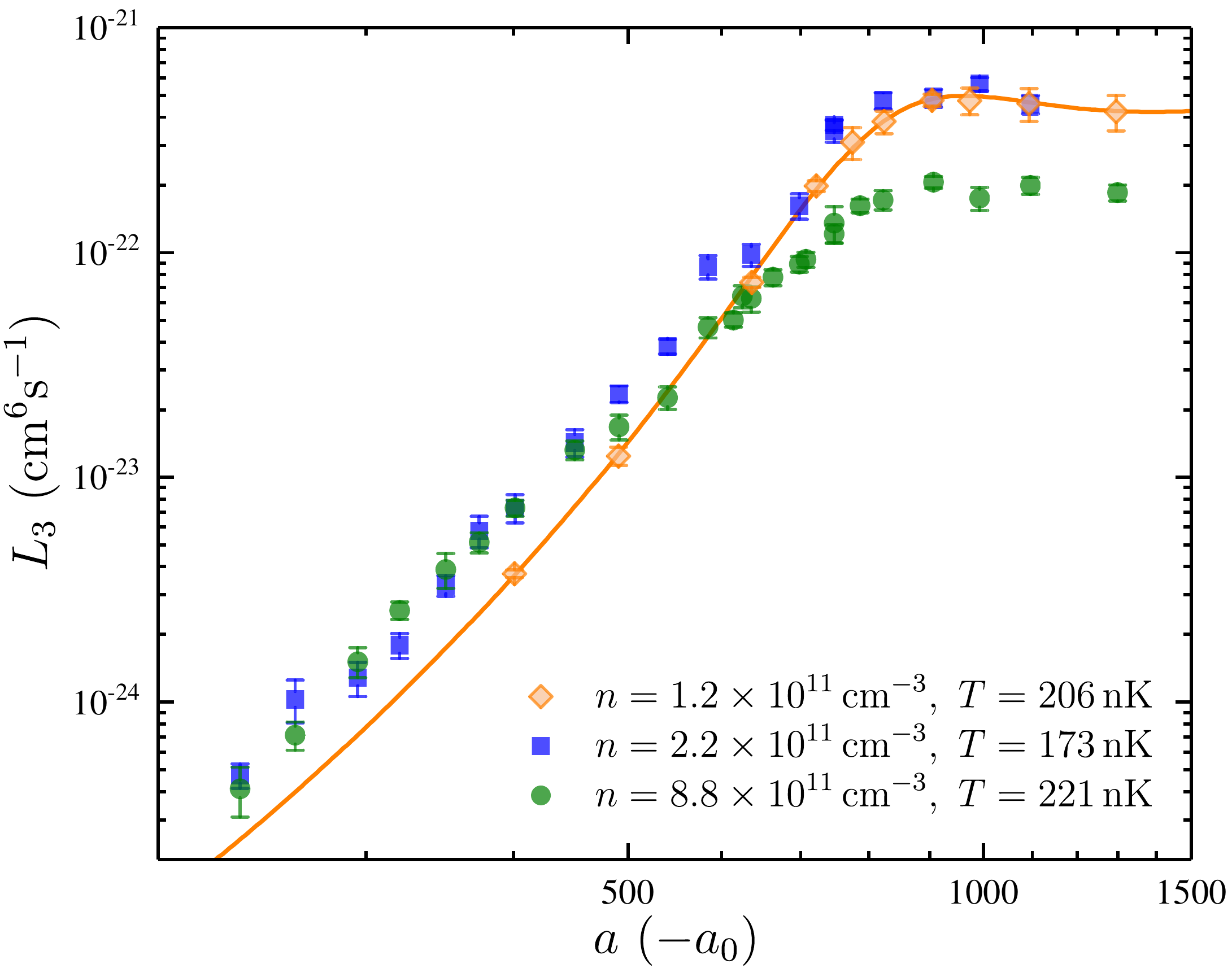} 
\caption{Suppression of the Efimov resonance in a high-density gas. Measurements of high- and intermediate-density samples performed with the same experimental conditions, contrasting only in the initial atom number. As a result, differential comparison of $L_{3}$ values between those two measurements is of greatest interest. Small $L_{3}$ deviations at low $\left|a\right|$ between the lowest-density data and other data are attributed to differing trap conditions that result in evaporation. However, for our highest-density data, we observed a strong suppression of $L_{3}$ near $a = a_{-}$.}
\label{fig:L3_vs_density_option1} 
\end{figure}

In addition to finite-temperature effects, we check the effect of high density on $L_{3}$ measurements. We prepare samples with varied densities yet similar temperatures of $\sim 200\thinspace$nK. Although measurements with the two lowest densities (where initial $\left|n a^{3} \right| = 1.3 \times 10^{-5}$ and $2.4 \times 10^{-5}$ at $a_{-}$) are consistent, we observe a suppression and shift of the Efimov resonance for our highest-density gas (see Fig.~\ref{fig:L3_vs_density_option1}), where $\left|n a^{3} \right| = 9.7 \times 10^{-5}$ at $a_{-}$ and the collision rate is no longer small as compared to the trapping frequency. The latter condition in particular can lead to systematic errors. A recently published study \cite{Wacker2018} on the same resonance as we discuss here reported counterintuitive temperature-induced shifts in the Efimov peak at high values of $\left|n a^{3}\right|$, the collision rate, and $n \lambda^{3}$. We see no such effects in the data (shown in Fig.~\ref{fig:L3_vs_T}) that we use to determine $a_{-}$; for those fits, we use only $\left|n a^{3}\right| < 4\times 10^{-5}$ and $n \lambda^{3} < 0.2$ \cite{SupplemMat}. The data shown in Fig.~\ref{fig:L3_vs_T} agree well with the prediction of Eq.~(\ref{eq:L3_T_dependent}): not just in the shape of $L_{3}(a, T)$ but in its overall amplitude $A$. The fact that, for all values of $T$, our fit $A$ is within $43\%$ of $1.0$ (which is consistent with small discrepancies in the density calibration) is further evidence that our results are not contaminated by high degeneracy, many-body effects, collisional avalanche, or misassignment of resonance peaks.

Our final value for $a_{-} = -908(11) \thinspace a_{0}$, plotted as a red star in Fig.~\ref{fig:aminus_vs_sres}, differs from the range of theoretical predictions \cite{Wang2012, Naidon2012, Naidon2017, Greene2017, DIncao2018} for the universal result of $a_{-} = -630 \pm 15\% \,a_0$ by many times our estimated error. How does this firmly established discrepancy compare to theoretical efforts to model the ``edges of universality''?

The range $a_{-} = [-11.2,-8.3]\, r_{\mathrm{vdW}}$ of theoretical predictions for the universal value arises because the calculated value of the ostensibly universal $a_{-}$ depends, even if only modestly, on the details of short-range treatment \cite{Wang2012}. It seems likely this variability will be only more pronounced for a regime where universality is already beginning to fail on its own. A key qualitative lesson from Ref.~\cite{Langmack2018} is the prediction of a nonuniversal value of $a_{-}\approx -12\, r_{\mathrm{vdW}}$ for $s_{\mathrm{res}} = 2.57$ and $a_{\mathrm{bg}} = -19.6 \, a_{0}$. However, going beyond the results from Ref.~\cite{Langmack2018}, we find that $a_{-}$ also depends on the number of bound states in the model for small $s_{\mathrm{res}}$ and $a_{\mathrm{bg}}$. In our theoretical effort to accurately describe three-body physics \cite{SupplemMat}, we constructed a more realistic multichannel model using a realistic hyperfine and Zeeman spin structure, with triplet and singlet scattering lengths constrained to equal our empirically determined values. The adjustable parameters are the inner walls of the van der Waals potentials tuned to give the desired number of bound states. The results are shown as black triangles in the inset of Fig.~\ref{fig:aminus_vs_sres}. We see that the predicted $a_{-}$ result more closely approximates our distinctly nonuniversal measurement as we go to a larger number of bound states. An empirical attempt to extrapolate to a very large number of bound states yields $a_{-}^{\mathrm{lim}} = -13.1 \, r_{\mathrm{vdW}}$ and $\eta^{\mathrm{lim}} = 0.21$. This is the first attempt to get a quantitatively accurate calculation for $\eta$ close to our measured value of $0.25(1)$. The reasonable agreement with the experimental value shows the importance of properly modeling the diatomic molecular spectra and its hyperfine structure \cite{Note2}. \footnotetext[2]{See Supplemental Material for further perspective on the progress of understanding interacting few-body systems, which includes Refs.~\cite{wang2011PRA, mestrom2017PRA, suno2002PRA, Stoof1988PRB, Tiesinga2000PRA, Berninger2013PRA, Lange2009PRA, Jachymski2013PRL, Hutson2007NJP, Nicholson2015PRA, Hutson2019ARX, Tanzi2018PRA, Gao1998PRA, Flambaum1999PRA, Gao2011PRA, Werner2012PRA, jonsell2004JPB}}

To conclude, we precisely measure dimer binding energies, the Feshbach resonance location, and the Efimov ground location. Our results (in particular, the observation of a definitively nonuniversal Efimov state location and its corresponding inelasticity parameter) suggest that more realistic models, like the one we used, can be necessary to fully understand and accurately describe few-body physics in ultracold atomic systems.

The authors thank Jeremy Hutson, John Bohn, Wilhelm Zwerger, and Richard Schmidt for useful discussions. This work is supported by NSF Physics \mbox{Frontier} Center Grant No.~1734006, NASA, the Marsico Research Chair, and the NIST.

\end{document}


\begin{abstract}
\end{abstract}
\title{Supplemental Material: Precision Test of the Limits to Universality in Few-Body Physics}

\author{Roman Chapurin}
\affiliation{JILA, National Institute of Standards and Technology, and the University of Colorado, Department of Physics, Boulder, Colorado 80309, USA}
\author{Xin Xie}
\affiliation{JILA, National Institute of Standards and Technology, and the University of Colorado, Department of Physics, Boulder, Colorado 80309, USA}
\author{Michael J. Van de Graaff}
\affiliation{JILA, National Institute of Standards and Technology, and the University of Colorado, Department of Physics, Boulder, Colorado 80309, USA}
\author{Jared S. Popowski}
\affiliation{JILA, National Institute of Standards and Technology, and the University of Colorado, Department of Physics, Boulder, Colorado 80309, USA}
\author{Jos\'e P. D'Incao}
\affiliation{JILA, National Institute of Standards and Technology, and the University of Colorado, Department of Physics, Boulder, Colorado 80309, USA}
\author{Paul S. Julienne}
\affiliation{Joint Quantum Institute, National Institute of Standards and Technology, and the University of Maryland, College Park, Maryland 20742, USA}
\author{Jun Ye}
\affiliation{JILA, National Institute of Standards and Technology, and the University of Colorado, Department of Physics, Boulder, Colorado 80309, USA}
\author{Eric A. Cornell}
\affiliation{JILA, National Institute of Standards and Technology, and the University of Colorado, Department of Physics, Boulder, Colorado 80309, USA}
\date{\today}
\maketitle
\section{Summary of the Synergism between Theory and Experiment}
Besides providing the details of some of the experimental and theoretical implementation of our studies, this supplementary material also provides the details on how theory and experiment worked together in order to produce highly accurate results. Our studies for the binding energies of Feshbach molecules (described in Section~\ref{sec:Eb_spectroscopy}) required a theoretical model consisting of a fully coupled channel approach \cite{chin2010} where realistic singlet and triplet Born-Oppenheimer potentials are used, supporting many dozens of $s$-wave molecular states, and still many more molecular states with non-vanishing orbital angular momenta. Such potentials were derived in Ref.~\cite{Falke2008} based on previous spectroscopic data allowing for the precise characterization of the singlet and triplet scattering lengths. Nevertheless, our current experimental data surpass that precision. Therefore, we further fine tune the potentials in order to fit our data and produce more refined values for the scattering lengths, the width and the position of the corresponding Feshbach resonance. The refinement procedure is described in Section~\ref{sec:Eb_spectroscopy}. 

Although our two-body model with realistic singlet and triplet potentials \cite{Falke2008} can describe our data for the binding energies quite accurately, this model is not suitable for performing three-body calculations due to the high number of molecular states the potentials can support, leading to a prohibitively large number of three-body channels to be described numerically. In our three-body approach \cite{wang2011PRA,Wang2012,wang2012PRLb,mestrom2017PRA,suno2002PRA} we determined the solutions of a system of two-dimensional partial differential equations which become highly oscillatory depending on the number of molecular states the potential can support, making it difficult to reach the necessary accuracy for our problem. In order to obtain a more numerically manageable model we replace the realistic singlet and triplet potentials by model potentials containing a much smaller number of bound states but still preserving the correct values for the singlet and triplet 
scattering lengths. Our three-body calculations use these reduced two-body models for different potentials containing a variable number of bound states in order to test their universality. This procedure is described in more detail in Section~\ref{sec: Efimov Resonance Location}. 

\section{Dimer Binding Energy Spectroscopy}
\label{sec:Eb_spectroscopy}

\subsection{Binding Energy Data}
\begin{figure}
\includegraphics[width=8.6cm,keepaspectratio]{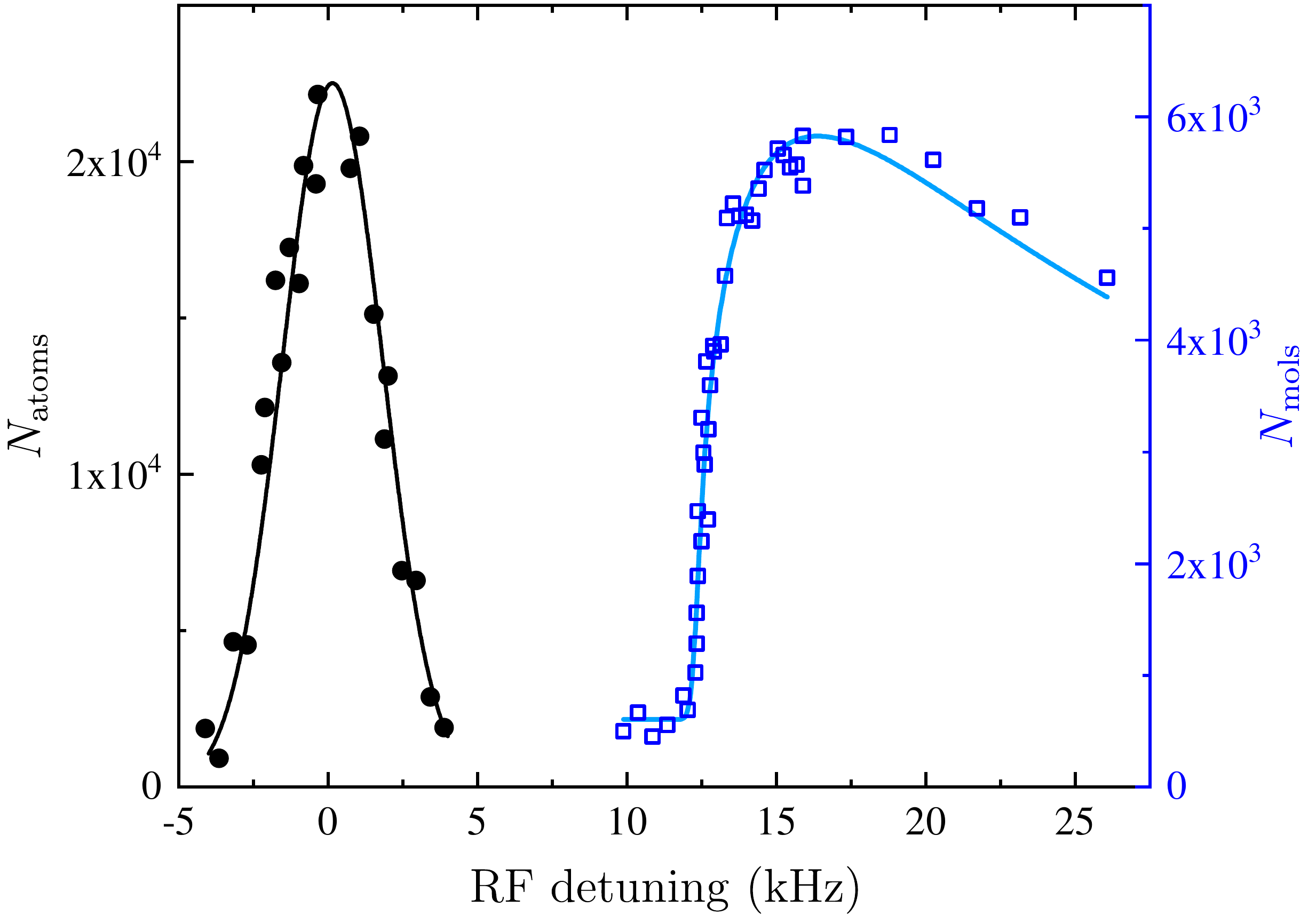} 
\caption{RF spectra of free-free transition (circles) and bound-free transition (squares) at $B=33.9575(3)\thinspace$G, where the RF frequency is detuned from $446.834413\thinspace$MHz. Values $N_{\mathrm{atoms}}$ and $N_{\mathrm{mols}}$ are proportional to the number of atoms and molecules detected during spectroscopy. The frequency difference between the atomic spectrum (black line) center and the onset of the dimer dissociation spectrum (blue line) corresponds to the dissociation threshold frequency in the confining trap.}
\label{fig:dimer_spectrum} 
\end{figure}

A compilation of the $E_{b}$ spectroscopy data is presented in Table~\ref{tab:Eb_data}. We take atomic spectra before and after each dimer dissociation measurement. The weighted mean $\bar{f}_{\mathrm{A}}$ of the two atomic lineshape centers ($f_{\mathrm{A,1}}$ and $f_{\mathrm{A,2}}$) defines the free-free transition frequency and, via the Breit-Rabi formula, the magnetic field $B$. To extract the bound-free dissociation threshold frequency, we subtract $\bar{f}_\mathrm{A}$ from the measured dimer dissociation spectrum and fit the spectrum to a function (see Fig.~\ref{fig:dimer_spectrum}) that is a convolution of the Franck-Condon factor \cite{Chin2005} and the Fourier-spectrum of the Gaussian-shaped RF dissociation pulse, whose duration, spanning $600\,\mu$s at high $E_b$ to $12\,$ms at low $E_b$, is chosen to be shorter than the dimer lifetime. The total uncertainty on the dissociation threshold frequency is taken as the fit error added in quadrature with uncertainty on $\bar{f}_{\mathrm{A}}$. Finally, we extract the free-space dimer binding energy $E_{b}$ by subtracting the total confinement-related frequency shift from the dissociation threshold frequency. We calculate the confinement shift for the final (free) and initial (bound) states and take their difference as the total confinement shift \cite{Idziaszek2006}. Due to relatively small trap frequencies of $\omega_{r}/2\pi = 28.64(66) \thinspace$Hz and $\omega_{z}/2\pi = 117.3(1.0) \thinspace$Hz and the final state being nearly non-interacting, the total confinement shift is equal, within uncertainty on our trapping frequencies, to the zero-point energy $87.3(1.4) \thinspace$Hz for all dissociation spectra.

\begin{table*}
\caption{Precise binding energy spectroscopy data. The $B$--field is determined from the weighted mean $\bar{f}_{\mathrm{A}}$ of two atomic lineshape centers $f_{\mathrm{A,1}}$ and $f_{\mathrm{A,2}}$, taken before and after each molecular dissociation spectrum. The frequency $f_{\mathrm{D}}$ corresponds to the dissociation threshold frequency in the confining trap, while $E_{b}$ corresponds to the free-space dimer binding energy.}
\label{tab:Eb_data}
\begin{ruledtabular} 
\begin{tabular}{llllll}
\multicolumn{1}{c}{free-free transition} & \multicolumn{1}{c}{free-free transition} & \multicolumn{1}{c}{free-free transition} & \multicolumn{1}{c}{magnetic field} & \multicolumn{1}{c}{confined dissociation} & \multicolumn{1}{c}{free-space} \\
\multicolumn{1}{c}{center $f_{\mathrm{A,1}}$}  & \multicolumn{1}{c}{center $f_{\mathrm{A,2}}$}  & \multicolumn{1}{c}{mean $\bar{f}_{\mathrm{A}}$} & \multicolumn{1}{c}{$B$} & \multicolumn{1}{c}{threshold $f_{\mathrm{D}}-\bar{f}_{\mathrm{A}}$} & \multicolumn{1}{c}{$E_{b}/h$} \\ 
\multicolumn{1}{c}{(MHz)} & \multicolumn{1}{c}{(MHz)} & \multicolumn{1}{c}{(MHz)} & \multicolumn{1}{c}{(G)} & \multicolumn{1}{c}{(kHz)} & \multicolumn{1}{c}{(kHz)} \\
\hline
446.870873(67) &	446.870945(63)	& 446.870911(52) & 33.7420(3) &	2.190(56) & 2.103(56) \\
446.861460(78) &	446.861312(76)	& 446.861384(76) & 33.7978(4) &	3.989(78) & 3.901(78) \\
446.852714(79) &	446.852546(82)	& 446.852634(82) &	33.8494(5) &	6.095(85) & 6.008(85) \\
446.834376(61) &	446.834445(57)	& 446.834413(48) &	33.9575(3) &	12.274(57) & 12.187(57) \\
446.826070(69) &	446.825957(59)	& 446.826004(60) &	34.0078(4) &	15.708(67) & 15.621(67) \\
446.816781(61) &	446.817086(55)	& 446.816950(116) &	34.0622(7) &	20.139(122) & 20.052(122) \\
446.800019(79) &	446.800145(76)	& 446.800085(71) &	34.1644(4) &	29.919(83) & 29.832(83) \\
446.783321(76) &	446.783547(73)	& 446.783439(96) &	34.2663(6) &	41.847(103) & 41.760(103) \\
446.748851(78) &	446.748911(73)	& 446.748883(57) &	34.4812(4) &	74.382(93) & 74.295(93) \\
446.731051(77) &	446.731033(101) & 446.731045(61) &	34.5940(4) &	95.395(137) & 95.307(137) \\
446.667723(82) &	446.667598(77)	& 446.667657(71) &	35.0060(5) &	200.293(406) & 200.205(406) \\
446.621332(86) &	446.621191(74)	& 446.621252(75) &	35.3198(5) &	308.716(436) & 308.628(436) \\
446.559027(89) &	446.558891(79)	& 446.558951(76) &	35.7593(6) &	508.003(582) & 507.916(582) \\
446.505076(79) &	446.505255(85)	& 446.505160(86) &	36.1582(7) &	742.262(1071) & 742.175(1071) \\
446.432524(82) &	446.432562(79) &	446.432544(58) &	36.7303(5) &	1167.324(1031) & 1167.237(1031) \\
\end{tabular}
\end{ruledtabular}
\end{table*}

\subsection{Two-Body Coupled-Channel Model}
We use coupled channels calculations~\cite{Stoof1988PRB,Tiesinga2000PRA} to calculate bound and scattering properties for two $^{39}$K atoms from the full two-body Hamiltonian:
\begin{align}
{H}=&-\frac{\hbar^2}{m} \nabla^2 + \sum_{SM_S} |SM_S \rangle V_S(r) \langle SM_S|\nonumber\\
&+ H_{hf}^{(1)}(B) + H_{hf}^{(2)}(B) \,,
\label{eq:2b_H}
\tag{S1}
\end{align}
where $m$ is the atomic mass and $r$ is the interatomic distance. In the above Hamiltonian, $V_{S=0}(\equiv V_S)$ and $V_{S=1}(\equiv V_T)$ are the respective electronic singlet and triplet Born-Oppenheimer molecular potentials for two interacting atoms, and $H_{hf}$ is the atomic hyperfine-Zeeman Hamiltonian in the presence of the external magnetic field $B$. The singlet and triplet potentials are given in Ref.~\cite{Falke2008}. In order to fine tune our interaction model with the experimental binding energy data, we have added a small correction $\Theta(r_{eq}^{i}-r) (r-r_{eq}^{i})^2\delta_i$ to each of the potentials of Ref.~\cite{Falke2008}, where $r_{eq}^{i}$ is the equilibrium position of the potential, and $\delta_i$ are fit parameters, labeled $\delta_S$ and $\delta_T$, used to fine-tune the singlet and triplet potentials. The resulting values are $\delta_S = 5.3196 \times 10^{-8}\,E_h$, $\delta_T = -2.4394 \times 10^{-9}\,E_h$, $r_{eq}^{S} = 7.3160\,a_0$, and $r_{eq}^{T} = 10.7371\,a_0$. See Section~\ref{subsec:fit_Eb_data} for a more detailed analysis of this fitting procedure.

We specifically consider the interactions of two $^{39}$K atoms with total spin projection $m_{F_1}+m_{F_2} = -2$, where $F$ is the atomic hyperfine quantum number and $m_F$ its azimuthal projection. This projection corresponds to the $|cc\rangle = |F_1 =1,m_{F_1} = -1 \rangle |F_2 =1,m_{F_2} = -1 \rangle |\ell,m_\ell\rangle$ spin channel relevant to our experiment, where $\ell = m_\ell=0$ for $s$-wave collisions at low collision energy. We find the magnetic dipole interaction terms~\cite{chin2010}, coupling $s$-waves to $d$-waves $(l=2)$ in Eq.~(\ref{eq:2b_H}), to be small and exclude them from the following discussion. Consequently the two-body radial Schr\"odinger equation can be written as
\begin{align}
\left[-\frac{\hbar^2}{m}\frac{d^2}{dr^2} + \frac{\hbar^2 \ell(\ell+1)}{mr^2}+\epsilon_\alpha \right ] f_\alpha(r)\nonumber\\
 +\sum_\beta V_{\alpha \beta}(r)f_\beta(r) = E f_\alpha(r) \,,
\label{eq:2b_H_radial}
\tag{S2}
\end{align}
where $\epsilon_\alpha$ is the channel energy for atoms in the hyperfine state $\alpha$ and $V_{\alpha \beta}$ are the corresponding interaction terms for the singlet and triplet potentials in the hyperfine basis. These two terms contain all the $B$ field dependence in the problem. Note that for s-wave ($l=0$) collisions, there are only five coupled spin channels (one open and four closed, all with $m_{F_1}+m_{F_2}=-2$), similar to the case with Na atoms in Ref.~\cite{Tiesinga2000PRA}. We thus include all $s$-wave spin channels in our calculations. Solutions of Eq.~(\ref{eq:2b_H_radial}) provide our results for the binding energy and scattering length for the $|F=1,m_F=-1\rangle$ $^{39}$K atoms used in this work.

\subsection{Extraction of Feshbach Resonance Parameters from $E_{b}$ Data}
\label{subsec:fit_Eb_data}

We adjust the coupled-channel model by performing a global fit to our $E_b(B)$ data. Specifically, we adjust the two fit parameters $\delta_S$ and $\delta_T$ that fine-tune the singlet and triplet potentials, respectively, and which ultimately determine the singlet and triplet scattering lengths $a_S$ and $a_T$. Since the predicted $E_b$ value at each magnetic field is predominately determined by a particular linear combination of $\delta_S$ and $\delta_T$, we perform the global fit in a rotated basis.

The fit allows us to constrain the corresponding linear combination of $a_S$ and $a_T$ to a high precision: $\sin(0.2496) a_S + \cos(0.2496) a_T = 1.926(2)\,a_0$. Additionally, we deduce the Feshbach resonance location $B_0 = 33.5820(14)\,$G, where the uncertainty is the fit error added in quadrature with $0.5\,$mG, the average uncertainty on $B$ in the $E_b(B)$ data. The $E_b = 1167\,$kHz data point leads to a significant increase in our reduced $\chi^2$ and we do not include it in our final fit. However, masking any data (single or multiple points) in our global fit results in the same $B_0$ within the quoted error.

We can combine our result with the constraint provided by the previously measured Feshbach resonance location at $560.72(20)\,$G~\cite{Roy2013} to extract $a_S = 138.85\,a_0$ and $a_T = -33.40\,a_0$, where the $a_S$ value is predominately determined by the resonance at $560.72(20)\,$G. These values are similar to the previously reported values $a_S = 138.49(12)\,a_0$ and $a_T = -33.48(18)\,a_0$ extracted from many Feshbach resonances~\cite{Falke2008,DErrico2007}.

This fine-tuned model determined by the $a_S$ and $a_T$ values predicts many two-body observables for any spin channel for all $B$, including $a(B)$, the effective range $r_e(B)$, $E_b(B)$, and the two-body inelastic rate coefficient $L_2(B)$. This information is also used to construct our reduced, but still full-spin model for three-body recombination, as described in Section~\ref{sec: Efimov Resonance Location}.

\begin{figure}[htbp]
\includegraphics[width=1.0\columnwidth,clip]{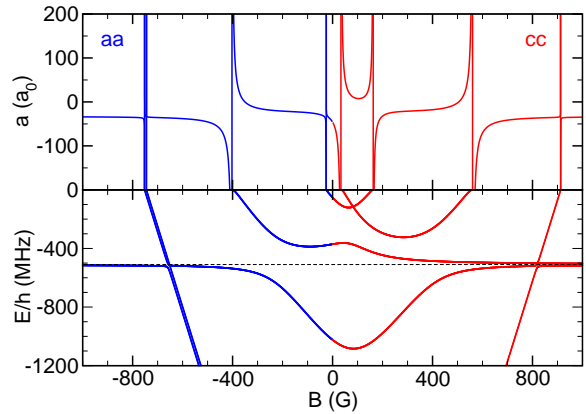}
\caption{Scattering length and $s$-wave bound state binding energies versus $B$ for channels $ii$ for $i=a,c$, where $E$ is relative to the channel energy $\epsilon_{ii}$ taken as 0. The dashed line shows where a pure van der Waals potential would have its last bound state with energy $E_{-1}$ for a scattering length of $a_T=-33.48\,a_0$ and the reduced mass of $^{39}$K atoms. }
\label{fig2S}
\end{figure}

Figure~\ref{fig2S} gives a broad overview of the two-body physics in $^{39}$K. The Figure shows the calculated scattering length and binding energies for the last several $s$-wave bound states for the two spin channels with $m_{F_1}+m_{F_2} = +2$ and $-2$. The former is the $|aa\rangle= |F_1 =1,m_{F_1} = +1 \rangle |F_2 =1,m_{F_2} = +1 \rangle |\ell=0,m_\ell=0\rangle$ spin channel and the latter the $|cc\rangle$ one, related by the reversed sign of the total spin projection. The Figure shows the results for the latter spin channel plotted for $-B$, since that is equivalent to switching the sign of the spin projection (as discussed, for example, in the case of Cs atoms by Berninger {\it et al.}~\cite{Berninger2013PRA}). In this way, both scattering length and the bound state spectrum are continuous across $B=0$.

There are several features that are evident from Fig.~\ref{fig2S}. One is the changing spin character of the near-threshold bound states. Thus, unlike many other alkali species, there is no bound state in the low $B$-field region with the same magnetic moment as the $aa$ or the $cc$ entrance channels, as evidenced by the lack of an $E(B)$ curve parallel with the $E=0$ line. At high fields, $|B| \gtrsim 500\,$G, the magnetic Zeeman splitting exceeds the hyperfine splitting and a bound state with magnetic moment similar to that of the entrance channels emerges. This high field state takes on dominant triplet character and lies where one would expect the last $s$-wave bound state in a van der Waals potential with the triplet scattering length (see the dashed line in Fig.~\ref{fig2S}).

Thus, the molecular states in the low field region where our experiment is performed do not have the entrance channel spin-character, but have a mixed character that is varying with $B$. This mixing is a consequence of the relatively small hyperfine coupling of $^{39}$K and the presence of several $s$-wave bound states that mix. The relatively strong $B$-dependence of the atomic and molecular states is one of the main reasons we needed to construct an all-spin representation of the Efimov physics in order to properly describe the amplitude of three-body recombination.

The second feature that is evident from Fig.~\ref{fig2S} is the presence of overlapping resonances. These can not be described simply by a single pole formula~\cite{Lange2009PRA,Jachymski2013PRL} of the conventional type,
\begin{equation}
\label{eq:avsB_1pole}
 a(B) = a_{\mathrm{bg}} \left ( 1 - \frac{\Delta}{B-B_0} \right ) \,,
 \tag{S3}
\end{equation}
where $a_{\mathrm{bg}}$ is the $B$-independent background scattering length in the absence of a resonance and $\Delta$ represents the width of the resonance. Their product $a_{\mathrm{bg}} \Delta$ represents the ``pole strength'' of the resonance. The need for a two-pole fit is already known for the case of $^{39}$K, since Roy {\it et al.}~\cite{Roy2013} had to use a two-pole formula to represent their coupled channels results for the strongly overlapping $59.92\,$G and $65.67\,$G resonance features in the $bb$ spin channel ($|b\rangle = |F=1,M_F=0\rangle$) of $^{39}$K. Consequently we fit our coupled channels $a(B)$ using a two-pole resonance expression similar to that of Ref.~\cite{Roy2013} (note: Ref.~\cite{Jachymski2013PRL} showed this sum form is equivalent to the product form in Ref.~\cite{Lange2009PRA}):
\begin{equation}
\label{eq:avsB_2pole}
a(B) = a_{\mathrm{bg}}^g \left ( 1 - \frac{\Delta_1}{B-B_{0,1}} - \frac{\Delta_2}{B-B_{0,2}} \right ) \,,
\tag{S4}
\end{equation}
where $a_{\mathrm{bg}}^g$ represents a ``global'' background for both resonances, which is not the same as that found from fitting a single pole. We obtain $a_{\mathrm{bg}}^g = $ $-19.599\,a_0$, widths $\Delta_1 = 54.772\,$G and $\Delta_2 = -36.321\,$G, $B_{0,1}=B_0$ and the second resonance location at $B_{0,2} = 162.341\,$G, consistent with the measured value $B_{0,2} = 162.35(18)\,$G from Roy {\it et al.}~\cite{Roy2013}. Rewriting Eq.~(\ref{eq:avsB_2pole}) in the one-pole form of Eq.~(\ref{eq:avsB_1pole}) gives
\begin{equation}
\label{eq:avsB_2pole_v2}
a(B) = a_{\mathrm{bg}}' \left ( 1 +\delta(B)- \frac{\Delta '}{B-B_0} \right ) \,,
\tag{S5}
\end{equation}
where $a_{\mathrm{bg}}'=a_{\mathrm{bg}} (1-\Delta_2/(B_{0,1}-B_{0,2})) =-14.070\,a_0$, $\Delta'=\Delta(1-\Delta_2/(B_{0,1}-B_{0,2}))^{-1} = 76.311\,$G, and $\delta(B)$ is a $B$-dependent correction that vanishes when $B=B_0$. A linear approximation for this case shows that $\delta(B) \approx -0.003052 (B-B_{0,1})$, giving rise to a weak $B$-dependent background away from the pole. Clearly, by construction, the pole strength of the local pole at $B_0$, $a_{\mathrm{bg}}^g \Delta_1 = a_{\mathrm{bg}}' \Delta'$ $=-1073.5\,a_0$G, is the same for either Eq.~(\ref{eq:avsB_2pole}) or~(\ref{eq:avsB_2pole_v2}).

Finally, it should be noted that there is an alternative method to calculate the ``pole strength'' for decaying resonances like those in the $cc$ channel which are coupled through the magnetic dipole interaction to exit channel $d$-waves and can decay to lower energy open channels of the Zeeman manifold. We can use either the formulation of Hutson~\cite{Hutson2007NJP} for the ``resonance length'' $a_{\mathrm{res}}$ of a decaying resonance with width $\delta$, or the equivalent formalism given for a decaying resonance by Nicholson {\it et al.}~\cite{Nicholson2015PRA}.

In the Hutson formulation, $a_{\mathrm{res}} (\delta/2) = a_{\mathrm{bg}} \Delta$~\cite{Note1}\footnotetext[1]{Note that the definition of $a_{\mathrm{res}}$ in Hutson~\cite{Hutson2007NJP} is twice the value defined by Eq.~(26) of Chin {\it et al.}~\cite{chin2010}}, and the maximum variation in $a(B)$ is $\pm a_{\mathrm{res}}/2$ when the magnitude of the detuning from the pole is $|B-B_0| = \delta/2$. Using the MOLSCAT scattering code of Ref.~\cite{Hutson2019ARX} and a basis set including $s$ and $d$ partial waves, we calculate for the $34\,$G pole $a_{\mathrm{res}} = -2.1281\times 10^7\,a_0$ and $\delta = 0.10084\,$mG, giving a pole strength of $-1073.0\,a_0$G; the same calculation determines ``local'' one-pole values of $a_{\mathrm{bg}}=-13.567\,a_0$ and $\Delta=78.975\,$G, giving the same product pole strength. This $a_{\mathrm{bg}}\Delta$ product is in good agreement with the value $-1073.5\,a_0$G calculated above from fitting a two-pole formula for $a(B)$ for the non-decaying resonances calculated with an $s$-wave basis only.

\subsection{Determining the $s_{\mathrm{res}}$ parameter}

Chin {\it et al.}~\cite{chin2010} defined a dimensionless parameter $s_{\mathrm{res}}$ to characterize Feshbach resonances when there is a long-range van der Waals potential. Although we never need to define this parameter to do our two-body coupled channels calculations or to set up and use our all-spin three-body calculations, we need to determine a value for it in making comparisons with other theories that make use of it.

As defined in Ref.~\cite{chin2010}, the definition of $s_{\mathrm{res}}$ presumed a two-channel model of an isolated single resonance in which a single bound state in a closed channel is coupled to an open entrance channel:
\begin{equation}
\label{eq:sres}
 s_{\mathrm{res}} = \frac{a_{\mathrm{bg}}\Delta}{\bar{a}} \frac{\delta\mu}{\bar{E}} \,,
 \tag{S6}
\end{equation}
where $\bar{a} = (4\pi/\Gamma(\frac14)]^2)\, r_{\mathrm{vdW}} \approx 0.955978\,r_{\mathrm{vdW}} \approx 61.77\,a_0$ \cite{Falke2008,Gribakin1993} is related to the van der Waals length, $r_{\mathrm{vdW}}= \frac12 \left ( m C_6/\hbar^2 \right )^\frac14\approx 64.61\,a_0$ for a van der Waals coefficient $C_6$ for the long-range potential; the corresponding energy scale is $\bar{E} = \hbar^2/(m \bar{a}^2)$. The open entrance channel has a background scattering length $a_{\mathrm{bg}}$, and the closed channel state differs by $\delta \mu$ in magnetic moment from the total magnetic moment of the two individual atoms. The ``bare" closed channel bound state is assumed to ramp linearly with slope $\delta \mu$ as the $B$ field is changed. The resonance coupling is characterized by the ``pole strength'' $a_{\mathrm{bg}} \Delta$ discussed in the last section. These approximations are adequate to characterize a wide variety of actual resonances involving various alkali-metal species.

\begin{figure}[htbp]
\includegraphics[width=\columnwidth,clip]{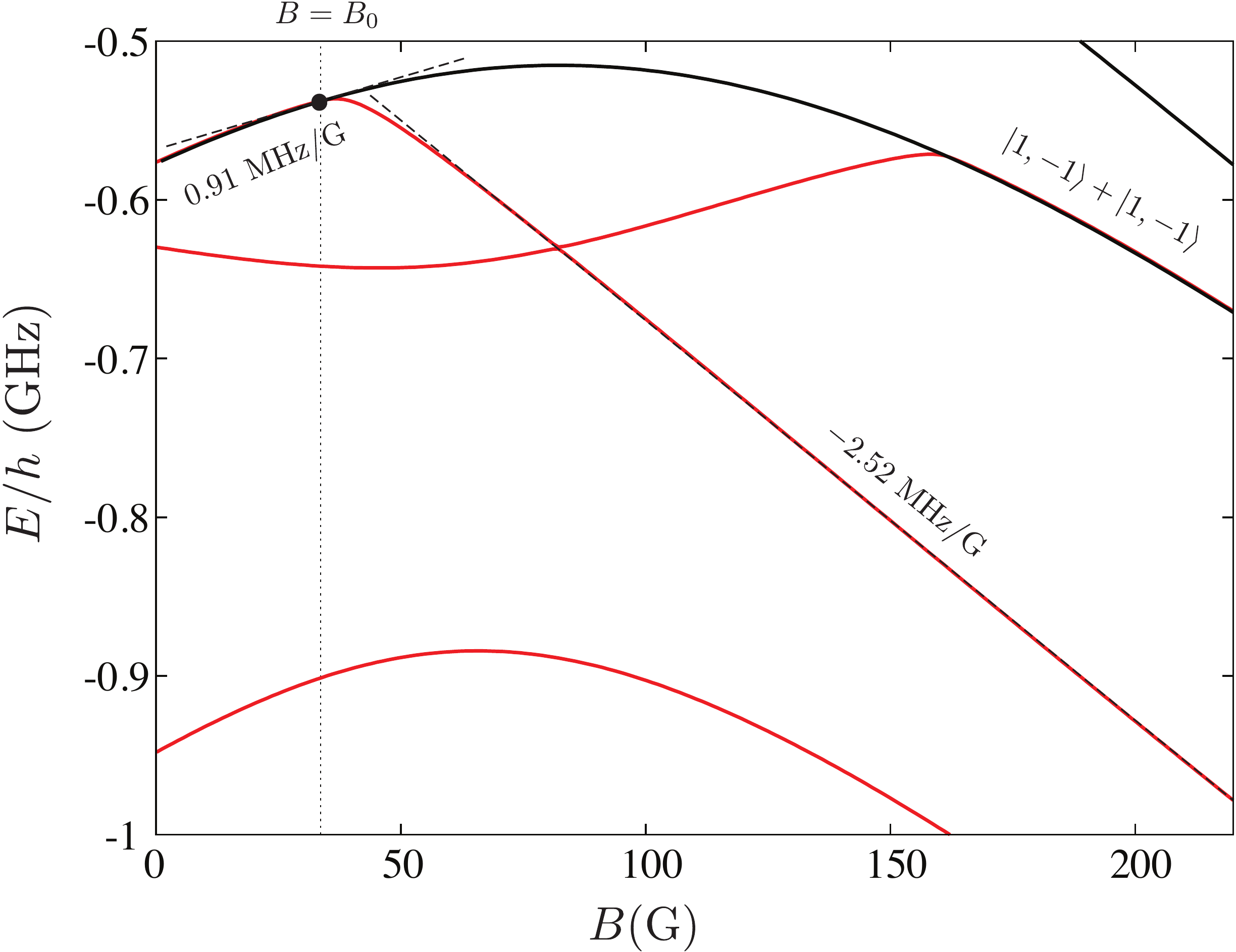}
\caption{Atomic (solid black) and calculated coupled channels bound state (solid red) absolute energies at different magnetic fields. The dashed lines with slopes of $0.91\,$MHz/G and $-2.52\,$MHz/G represent the atomic and molecular magnetic moments near and away from $B_0$ respectively; their difference is $\delta\mu/h\approx 3.43\,$MHz/G.}
\label{fig3S}
\end{figure}

The parameters $a_{\mathrm{bg}}$, $\Delta$, and $\delta \mu$ defining $s_{\mathrm{res}}$ in the model of Ref.~\cite{chin2010} were assumed to be independent of $B$-field tuning near the resonance. Thus, we will first attempt to estimate a plausible value for $\delta \mu$ using our full two-body coupled channels calculations of the bound states. Figure~\ref{fig3S} shows that a straight line with a slope of $\delta \mu/h =-2.52\,$MHz/G provides a good approximation to the bound state energy away from threshold to represent a ramping closed channel state. Using this value of the slope and subtracting it from the the slope of $0.91\,$MHz/G for the atomic threshold near the $34\,$G resonance (dashed line in Fig.~\ref{fig3S}), along with the previously determined $a_{\mathrm{bg}}\Delta$ product determines a value of $s_{\mathrm{res}}\approx2.45$. We note that the atomic magnetic moment near the $34\,$G resonance is still dependent on the $B$ field, which introduces some ambiguity on the value of $s_{\mathrm{res}}$ as defined in Eq.~(\ref{eq:sres}). In our case, the value of $0.91\,$MHz/G is obtained by calculating the slope at $B=B_0$. 

There are various approaches to extract the $s_{\mathrm{res}}$ value when $a_{\mathrm{bg}}$, the spin character or the magnetic moment varies across a resonance region. We follow the scheme based on the effective range similar to that discussed by Roy {\it et al.}~\cite{Roy2013} (in their Supplemental Material) and Tanzi {\it et al.}~\cite{Tanzi2018PRA}. We calculate the effective range $r_e$ from coupled channels calculations and use it to determine a value for $s_{\mathrm{res}}$. The effective range is the parameter that characterizes how the scattering properties vary with energy away from threshold. There is an analytic formula for $r_e$ for a two-channel representation of a local resonance in a van der Waals potential. It includes a term for a single channel with a van der Waals potential and a scattering length $a$, having the analytic form \cite{Gao1998PRA, Flambaum1999PRA}
\begin{equation}
\label{eq:re_1channel}
r_e^{(1)}(a) = \frac{\Gamma(\frac14)^4}{6\pi^2} \bar{a} \left ( 1 - 2\frac{\bar{a}}{a} + \left (\frac{\bar{a}}{a} \right )^2 \right ) \,,
\tag{S7}
\end{equation}
where the numerical factor $\Gamma(\frac14)^4/(6\pi^2)\approx 2.91791$. The full two-channel expression also includes a resonant term depending on $s_{\mathrm{res}}$ and is given by Gao~\cite{Gao2011PRA}; also cited by Werner and Castin~\cite{Werner2012PRA} as their Eq.~(185), and is used in Refs.~\cite{Roy2013,Tanzi2018PRA}:
\begin{equation}
\label{eq:re_2channel}
 r_e(B) = -2R^* \left ( 1- \frac{a_{\mathrm{bg}}}{a(B)} \right )^2 + r_e^{(1)}(a(B)) \,,
 \tag{S8}
\end{equation}
where the parameter $R^* = \bar{a}/s_{\mathrm{res}}$. Using Eq.~(\ref{eq:re_2channel}) allows a value of $s_{\mathrm{res}}$ to be found from the knowledge of $r_e$ at the resonance pole position, where $a \to \infty$, and $r_e(B_0) = -2\bar{a}/s_{\mathrm{res}}+ 2.91791\bar{a}$. Solving this latter expression for $s_{\mathrm{res}}$ given $r_e$ is trivial. Using our calculated value of $r_e(B_0) \approx 133.91\,a_0$ at the $34\,$G resonance pole we obtain $s_{\mathrm{res}}\approx 2.68$ via this method.
\begin{figure}[htbp]
\includegraphics[width=\columnwidth,clip]{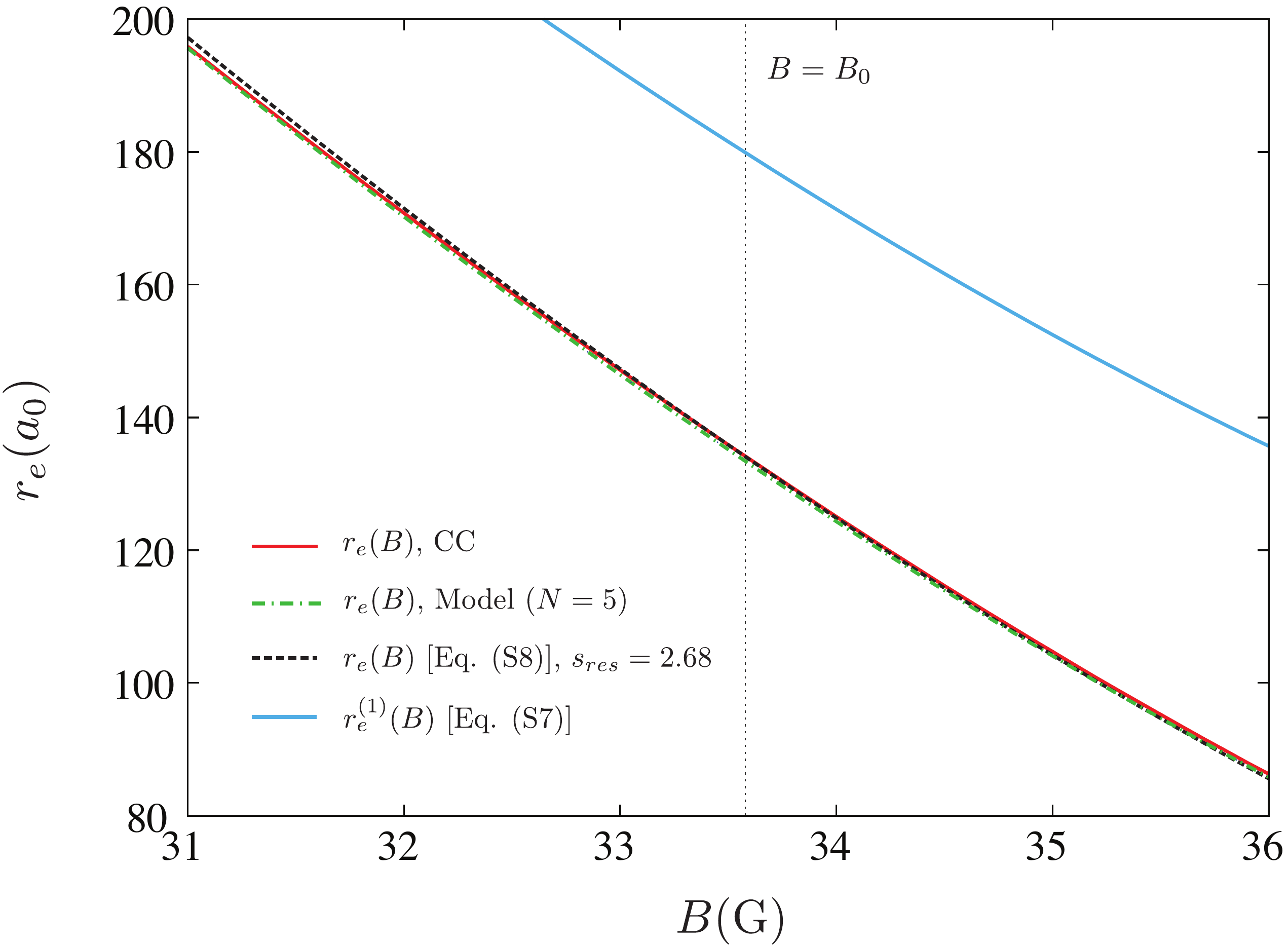}
\caption{Various calculations of the effective range $r_e$ versus $B$ for the $cc$ channel near the pole. The solid blue line is the single channel formula, Eq.~(\ref{eq:re_1channel}). The three nearly indistinguishable lines are: the solid red line for the full coupled channels (CC) result; the dashed line from formula Eq.~(\ref{eq:re_2channel}) for $s_{\mathrm{res}}\approx2.68$; and the dot-dashed line calculated with our two-body reduced model described in Section~\ref{subsec:3b_model}}
\label{fig4S}
\end{figure}

Figure~\ref{fig4S} shows that the exact two-channel Eq.~(\ref{eq:re_2channel}) gives an excellent approximation to the true coupled channels result for $r_e(B)$ when $s_{\mathrm{res}} \approx\:2.68$. Figure~\ref{fig4S} also shows that the values of $r_e(B)$ determined by our reduced two-body model (with $N=5$ singlet bound states), used in our three-body calculations and described in Section~\ref{subsec:3b_model}, are also nearly indistinguishable from the coupled channels calculations. We note that the value $s_{\mathrm{res}}\approx 2.68$ is also consistent with the $s_{\mathrm{res}}\approx 2.45$ obtained from the analysis of the magnetic moment $\delta\mu$. Therefore, we will take the arithmetic mean 2.57 as the best value for $s_{\mathrm{res}}$.

Finally, we comment that it can be a useful heuristic to think of $s_{\mathrm{res}}$ as defining the width of the Feshbach resonance's universal region. For sufficiently small positive magnetic field detunings $\delta_B = B - B_0$, the bound state is fully in the open channel, where the binding energy and the scattering length follow the universal expressions $E_b \approx \hbar^2/(ma^2)$ and $a(B) \approx a_{\mathrm{bg}}\Delta/\delta_B$, such that $\partial E_b/\partial \delta_B \approx 2(\hbar^2/m)\delta_B/(a_{\mathrm{bg}}\Delta)^2$. At large values of $\delta_B$, the bound state asymptotes to the closed channel and $\partial E_b/\partial \delta_B \approx \delta \mu$. We can define a crossover scattering length $a_c$ for which the open-channel prediction for $\partial E_b/\partial \delta_B $ is half of the ultimate asymptotic limit. For $s_{\mathrm{res}} \lesssim 1$ and $|a_{\mathrm{bg}}|\lesssim r_{\mathrm{vdW}}$, we might then expect more open-channel, universal behavior when $a \gg a_c \approx 4 r_{\mathrm{vdW}}/s_{\mathrm{res}}$.
\begin{table*} 
\caption{Efimov resonance measurement conditions and fit results. We extract the Efimov peak location, width and height from the zero-temperature Eq.~(1), limiting fits to data points for which $\left|a\right| < \lambda/10$. We extract $a_{-}$, $\eta$ and $A$ from the finite-temperature Eq.~(2).}
\begin{ruledtabular}
\begin{tabular}{llllllllll}
$\bar{T}$ & $T_{\mathrm{i}}$ & $\langle n \rangle _{\mathrm{i}}$ & $\bar{\omega}/2\pi$ & peak location & peak width & peak height & $a_{-}$ & $\eta$ & A \\
nK & nK & $10^{11} \thinspace \mathrm{cm^{-3}}$ & Hz & $-a_{0}$ & & & $-a_{0}$ & & \\
\hline
$30$ & $27$ & $0.40$ & $33.7$ & $891(8)$ & $0.22(1)$ & 1.29(4) & 910(9) & 0.24(1) & 1.40(4) \\
$53$ & $46$ & $0.70$ & $40.8$ & $856(10)$ & $0.22(1)$ & 1.28(5) & 904(12) & 0.24(1) & 1.43(6) \\
$94$ & $85$ & $1.07$ & $51.4$ & $829(7)$ & $0.24(1)$ & 0.92(3) & 890(7) & 0.24(1) & 1.08(2) \\
$206$ & $186$ & $1.21$ & $65.7$ & $807(14)$ & $0.30(2)$ & 0.65(3) & 921(6) & 0.25(1) & 0.91(1) \\
$510$ & $407$ & $3.52$ & $89.9$ & $811(46)$ & $0.44(3)$ & 0.54(5) & 917(9) & 0.26(1) & 0.89(1) \\
\hline
$173$ & $160$ & $2.18$ & $51.1$ & & & & & & \\
$221$ & $194$ & $8.76$ & $51.1$ & & & & & & 
\label{table:Efimov_table}
\end{tabular}
\end{ruledtabular}
\end{table*}

\section{Efimov Resonance Location} 
\label{sec: Efimov Resonance Location}

\subsection{Data Conditions and Fit Results}
A compilation of Efimov resonance measurement conditions and fit results is presented in Table~\ref{table:Efimov_table}. Each temperature data set is described by a time-averaged temperature $\bar{T}$, initial temperature $T_{\mathrm{i}}$, initial mean density $\langle n \rangle _{\mathrm{i}}$ and a mean trap frequency $\bar{\omega}$. Atom number and temperature at each interrogation time are extracted from absorption images of the sample after a fixed time-of-flight, assuming Maxwell-Boltzmann distribution of the sample. Trapping frequencies are measured using the slosh motion of the sample along two orthogonal directions. Efimov $L_{3}/a^{4}$ peak locations, widths and heights are determined from fits to the zero-temperature zero-range expression (Eq.~(1) of the main text), limiting fits to data points for which $\left|a\right| < \lambda/10$ (thermal wavelength) and redefining $a_{-}$ as the peak location, $\eta$ as the peak width and the fit prefactor as the peak height. We extract the true Efimov ground state location $a_{-}$ and $\eta$ from a fit to the finite-temperature zero-range expression (Eq.~(2) of the main text), where the fit amplitude prefactor deviation from unity describes the uncertainty in our absolute density calibration. We do not include fit results for high $n a^{3}$ data, for which there appears to be some unaccounted for many-body effect. 

\subsection{Three-body Coupled-Channel Model}
\label{subsec:3b_model}
Our three-body calculations for $^{39}$K atoms were performed using the adiabatic hyperspherical representation \cite{suno2002PRA, wang2011PRA,Wang2012,wang2012PRLb,mestrom2017PRA}, with atoms containing the proper hyperfine structure. In order to incorporate such effects we have used Feshbach projectors in an approach similar to the one used in Ref.~\cite{jonsell2004JPB}. In the hyperspherical representation the hyperradius $R$ determines the overall size of the system, while all other degrees of freedom are represented by a set of hyperangles $\Omega$. Within this framework, the three-body adiabatic potentials $U$ and channel functions $\Phi$ are determined from the solutions of the hyperangular adiabatic equation:
\begin{align}
&\left[\frac{\Lambda^2(\Omega)+15/4}{2\mu R^2}\hbar^2+\varepsilon_{\alpha}\right]\Phi_{\alpha}(R;\Omega)\nonumber\\
&+\sum_{\beta}\sum_{i<j}V_{\alpha\beta}(r_{ij})\Phi_{\beta}(R;\Omega)=U(R)\Phi_{\alpha}(R;\Omega) \,.
\tag{S9}
\end{align}
This expression contains the hyperangular part of the kinetic energy, expressed through the grand-angular momentum operator $\Lambda^2$ and the three-body reduced mass $\mu=m/\sqrt{3}$. In our formulation, as well as that of Ref.~\cite{jonsell2004JPB}, the multichannel structure of interatomic interactions uses the same two-body interaction potentials as in Eq.~\ref{eq:2b_H_radial}. The resonant spin channels we consider are fully symmetric with $m_{F_1}+m_{F_2}+m_{F_3}=-3$, with at least one pair, $ij$, with $m_{F_i}+m_{F_j}=-2$ and the third, $k$, atom in the $|F_k=1,m_{F_k}=-1\rangle$ state. We note that, since our interaction model incorporates the proper hyperfine structure, as well as singlet and triplet interactions, the correct values of all two-body resonance parameters, including $s_{\mathrm{res}}$ and $a_{\mathrm{bg}}$, are naturally built in.

For our three-body calculations near the $B_0=33.5820\,$G resonance, we have replaced the actual singlet and triplet potentials from Ref.~\cite{Falke2008} by two Lennard-Jones potentials, $v_{S}(r)=-C_6/r^6(1-\lambda_{S}^6/r^6)$, and $v_{T}(r)=-C_6/r^6(1-\lambda_{T}^6/r^6)$, with $\lambda_{S}$ and $\lambda_{T}$ adjusted to correctly produce the singlet and triplet scattering lengths but with a much smaller number of bound states than the real interactions. In practice, we allowed in our model for small variations in $a_{S}$ and $a_{T}$ to fine tune the background scattering length $a_{\mathrm{bg}}$ and width of the Feshbach resonance, $\Delta $, thus ensuring the proper pole strength $a_{\mathrm{bg}}\Delta$. The fact that we are replacing the realistic singlet and triplet potentials but preserving the atomic hyperfine interactions eliminates the necessity to specify the value of $s_{\mathrm{res}}$, while still accurately describing the properties of the Feshbach resonance (see Fig.~\ref{fig4S}). Moreover, our reduced model has the correct mixing of the spin states and thus properly describes the relevant molecular channels for three-body recombination. These aspects, absent in most previous three-body models \cite{Schmidt2012,Wang2014,Langmack2018}, are important for the low $B$-field Feshbach resonances in $^{39}$K, where the $B$-field dependence of the hyperfine interactions leads to a strongly mixed spin character of both atomic and molecular states. 

The value for the Efimov resonance position $a_{-}$ is obtained using $v_{S}$ and $v_{T}$ potentials supporting different numbers of $s$-wave bound states. This is done by solving the hyperradial Schr\"odinger equation \cite{wang2011PRA},
\begin{align}
&\left[-\frac{\hbar^2}{2\mu}\frac{d^2}{dR^2}+U_{\nu}(R)\right]F_{\nu}(R)\nonumber\\
&~~~~~~+\sum_{\nu'}W_{\nu\nu'}(R)F_{\nu'}(R)=EF_{\nu}(R) \,,
\label{Schro}
\tag{S10}
\end{align}
where $\nu$ is an index that labels all necessary quantum numbers to characterize each channel, and $E$ is the total energy. From the above equation we determine the scattering $S$-matrix, and the resulting recombination rate $L_3$. The value for $a_-$, as well as the inelasticity parameter $\eta$, are then determined via fitting to the universal formula Eq.~(1) of the main text. The predicted values for $a_{-}$ and $\eta$ are listed in Table~\ref{Tab3BP} for different numbers of $s$-wave singlet bound states $N_{S}\equiv N$ our model potential can support. The number of triplet $s$-wave states is given by $N_{T}=N-1$. We also list the total number of bound states in Table~\ref{Tab3BP}, including all partial waves, after adding the hyperfine interactions that cause the mixing between singlet and triplet states.

From Table~\ref{Tab3BP} we see a considerable dependence of $a_{-}$ on the number of $s$-wave states approaching a limiting value that differs only $8\%$ from the experimental finding of $a_{-}=-14.05(17) \, r_{\rm vdW}$. This stronger dependence on the number of bound states is in contrast to the results obtained for broad resonances \cite{Wang2012}. We also see a similar behavior for the inelasticity parameter $\eta$, whose limiting value is 0.21. This remarkable level of agreement for $\eta$ indicates that our model is capable of properly describing the reaction rates in the system. Given our assumption that the pairwise additive long-range potentials are sufficient to account for the Efimov physics, our three-body model should be valid for $B$-fields in which the two-body bound states and scattering properties are described properly. Possible future improvements in our three-body model would include the effects of the magnetic-dipole interactions that couple $s$- and $d$-wave two-body interactions, a more realistic model of electronic exchange interactions, and checking the effect of non-additive short-range three-body potentials.
\begin{table}
\caption{Values for the three-body parameter $a_{-}$ and inelasticity parameter $\eta$ as a function of singlet number of $s$-wave bound states $N_{S}\equiv N$ (or the total number of bound states $N_{\mathrm{tot}}$ including those of nonvanishing rotational angular momenta). We extrapolate to high-$N$ by fitting to different functional forms and take the fit results spread as the uncertainty for the limiting $a_{-}$ and $\eta$ values. The last row lists experimental results obtained in the present work.}
\vspace{1pt}
\begin{ruledtabular}
\begin{tabular}{llll}
$N$ & $N_{\mathrm{tot}}$ & $a_{-}/r_{\rm vdW}$ & $\eta$\\ \hline
2 & 6  & $-7.61$ & 0.10\\
3 & 27 & $-11.20$ & 0.19 \\
4 & 59 & $-12.27$ & 0.20 \\
5 & 105 & $-12.69$ & 0.21 \\
$\vdots$ & $\vdots$ &$ \vdots$ & $\vdots$\\
$\infty$ & $\infty$ & $-13.1(3)$ & 0.21(1) \\
\hline
Exp. & & $-14.05(17)$ & 0.25(1)
\end{tabular}
\end{ruledtabular}
\label{Tab3BP}
\end{table}